# Gini-Type Index for Ageing/Rejuvenating Populations

Mark. P. Kaminskiy

Abstract -- Application of some basic notions and statistics of ageing distributions used in mathematical theory of reliability including the Gini-type index is discussed as a methodological tool for investigation of human population ageing and longevity. In opposite to the traditionally used median and mean lifetime, the GTI requires only one calendar year snap-shot data to reveal the population ageing or rejuvenating. The case study illustrating the suggested techniques is based on the Australia 1921 and Australia 2009 mortality data from the Human Mortality Database (http://www.mortality.org)

## I. INTRODUCTION

Increasing of population longevity, being beneficial for individuals and society as a whole, is a problem, which has become more and more important across the world in recent years not only for the governments with social security systems, but for insurance companies and funds providing pension benefits as well [1, 2].

From statistical standpoint, aging reveals itself, when the median *age of a country* increases with time. Ageing can also be observed as increase of lifetime (say, mean lifetime) of a population. Thus, in investigating ageing, one can deal with two types of distributions: *the distribution of population age,* and the *distribution of lifetime* of a given population (mortality data). It should be noted, that in both cases, one has to compare two respective statistics calculated at different years. For example, if the median $M(T_1)$ is the median age in $T_1$ year, and $M(T_2)$ is the median age in $T_2$ year, $(T_2 > T_1)$, we can talk about ageing if $M(T_2) > M(T_1)$ $T_1 < T_2$.

In this study, we will consider ageing data analysis applied to human mortality data. The main question we are going to address below is to what extent the notions and statistics of ageing distributions used in mathematical theory of reliability [3] can be applied to investigation of human population longevity. We also suggest applying the reliability theory and index of ageing/rejuvenating [4, 5] to the mortality data. This index requires only one calendar year snapshot data to reveal whether the population is ageing or rejuvenating.

## II. AGEING and REJUVENATING DISTRIBUTIONS

Let's have a continuous lifetime distribution with a probability density function $f(t)$, where $t$ is the lifetime (life duration). The respective *survival* (*survivor*, or *reliability*) function $S(t)$ is introduced as

$$S(t) = 1 - \int_0^t f(\tau)d\tau, \tag{1}$$

The corresponding *rate* (or *force*) *of mortality* (*death rate*, also *hazard* or *failure rate* in reliability) is the lifetime conditional probability density function, i.e.,

$$h(t) = \frac{f(t)}{S(t)} \tag{2}$$

A population is called *ageing* if its rate of mortality is increasing. Similarly, a population is called *rejuvenating* if its rate of mortality is decreasing. It can be shown [3] that the only distribution having a constant mortality rate is the exponential distribution. A population having a constant mortality rate is non-aging. We can also state that the exponential distribution is the boundary distribution between the class of aging distributions and the class of rejuvenating distributions.

## III. GINI-TYPE INDEX for AGING/REJUVENATING DISTRIBUTIONS

The introduced in [4, 5] Gini-type index (GTI) shows how different a given ageing (or rejuvenating) distribution is from the constant mortality rate (non-ageing) distribution i.e., the exponential distribution. To an extent, the GTI is similar to the *Gini index* used in macroeconomics for comparing an income distribution of a given country with the uniform distribution covering the same income interval. The Gini index is used as a measure of income inequality [6]. The index takes on the values between 0 and 1. The closer the index value to zero, the closer the distribution of interest is to the uniform one. The interested reader could find the index values sorted by countries in [7]. It includes the United Nations and CIA data.

The GTI takes on the values between –1 and 1. The closer the index's value is to zero, the closer the distribution of interest is to the exponential distribution. A positive value of the index indicates an ageing distribution. Respectively, a negative value of the GTI indicates a rejuvenating distribution. The GTI index is introduced as follows.

Let's have a population, whose lifetime distribution belongs to the class of the ageing distributions with a rate of mortality $h(t)$ at lifetime $t$. The respective *cumulative rate of mortality* is

$$H(t) = \int_0^t h(\tau)d\tau \qquad (3)$$

and is concave upward [3,4].

Consider a time interval $[0, T]$. The cumulative mortality rate function at lifetime $T$ is $H(T)$, and the respective survival function is $S(T)$. Now, introduce $h_{eff}$, as the rate of mortality of the exponential distribution with the survival function $S(t)$ equal to the survival function of interest at the time $t = T$, i.e.,

$$h_{eff}(T) = -\frac{\ln(S(T))}{T} \qquad (4)$$

where $S(T) = \exp(-h_{eff}T)$.

In other words, the introduced exponential distribution with parameter $h_{eff}$, at $t = T$, has the same value of the cumulative rate of mortality as the ageing distribution of interest, as depicted in Figure 1 below.

The GTI index is then introduced as

$$GTI(T) = 1 - \frac{\int_0^T H(t)dt}{0.5Th_{eff}(T)T} = 1 - \frac{2\int_0^T H(t)dt}{TH(T)} = 1 - \frac{2\int_0^T \ln(R(t))dt}{T\ln(R(T))} \qquad (5)$$

In terms of Figure 1, the index is defined as one minus the ratio of areas $A$ and $A + B$. It is easy to check that the above expression also holds for the rejuvenating (decreasing mortality rate) distributions, for which $H(t)$ is concave downward.

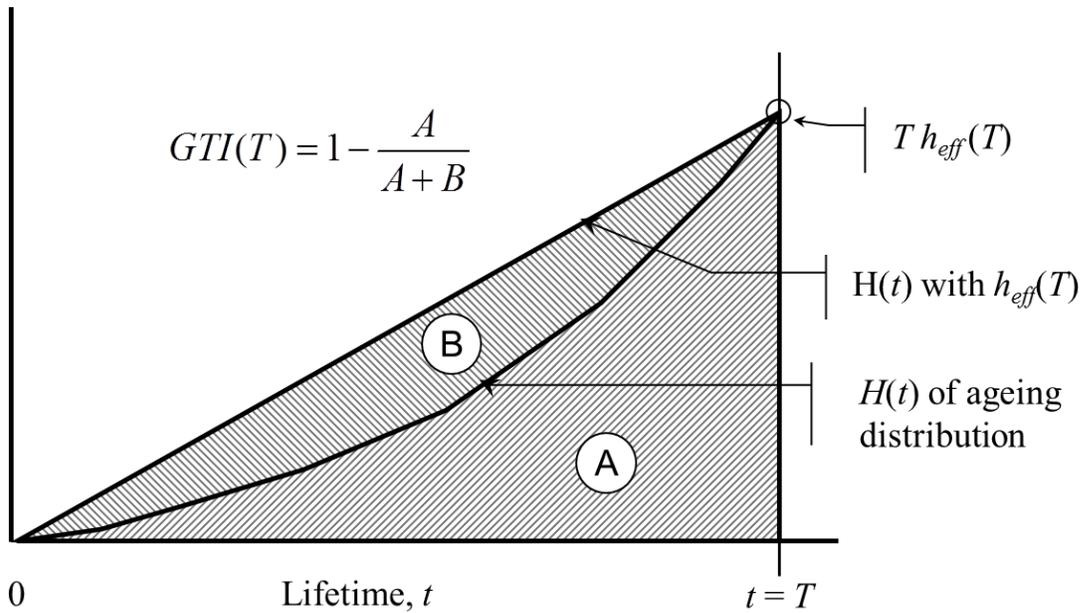

Figure 1. Graphical interpretation of the Gini-Type index for an aging distribution

It is clear that the GTI satisfies the following inequality: $-1 < C(T) < 1$. The index is positive for the aging distributions, negative for the rejuvenating distributions and it is equal to zero for the constant mortality rate distribution, i.e., the exponential distribution. Note that the suggested index is, in a sense, distribution–free, i.e. it is applicable to any continuous distribution used as a lifetime distribution.

For some distributions, the GTI can be expressed in a closed form. For example, in the case of the Weibull distribution with shape parameter $\beta$ the GTI index can be expressed as

$$C = 1 - \frac{2}{\beta + 1} \qquad (6)$$

It is worth noting that, in this case, the GTI depends neither on the scale parameter of the distribution, nor on time interval *T*. Interestingly, $C(\beta) = -C\left(\frac{1}{\beta}\right)$, which is illustrated by Table 1.

*Table 1. Gini-Type index for Weibull Distribution as Function of Shape Parameter $\beta$*

| Shape Parameter $\beta$ | GT Index | Lifetime Distribution |
|---|---|---|
| 5 | 0.6(6) | Ageing |
| 4 | 0.6 | Ageing |
| 3 | 0.5 | Ageing |
| 2 | 0.3(3) | Ageing |
| 1 | 0 | Constant mortality rate |
| 0.5 | –0.3(3) | Rejuvenating |
| 0.3 | –0.5 | Rejuvenating |
| 0.25 | –0.6 | Rejuvenating |
| 0.2 | –0.6(6) | Rejuvenating |

IV. CASE STUDY

The data used in our case study are from the Human Mortality Database (HMD) that "was created to provide detailed mortality and population data to researchers, students, journalists, policy analysts, and others interested in the history of human longevity. The project began as an outgrowth of earlier projects in the Department of Demography at the University of California, Berkeley, USA, and at the Max Planck Institute for Demographic Research in Rostock, Germany "[8].

The two sets of data, which were analyzed, include Australia 1921, and Australia 2005. The data are illustrated in Figure 2. The results of evaluation of the GTI given by Equation (5) and the survival function (1) are given in Table 2 below.

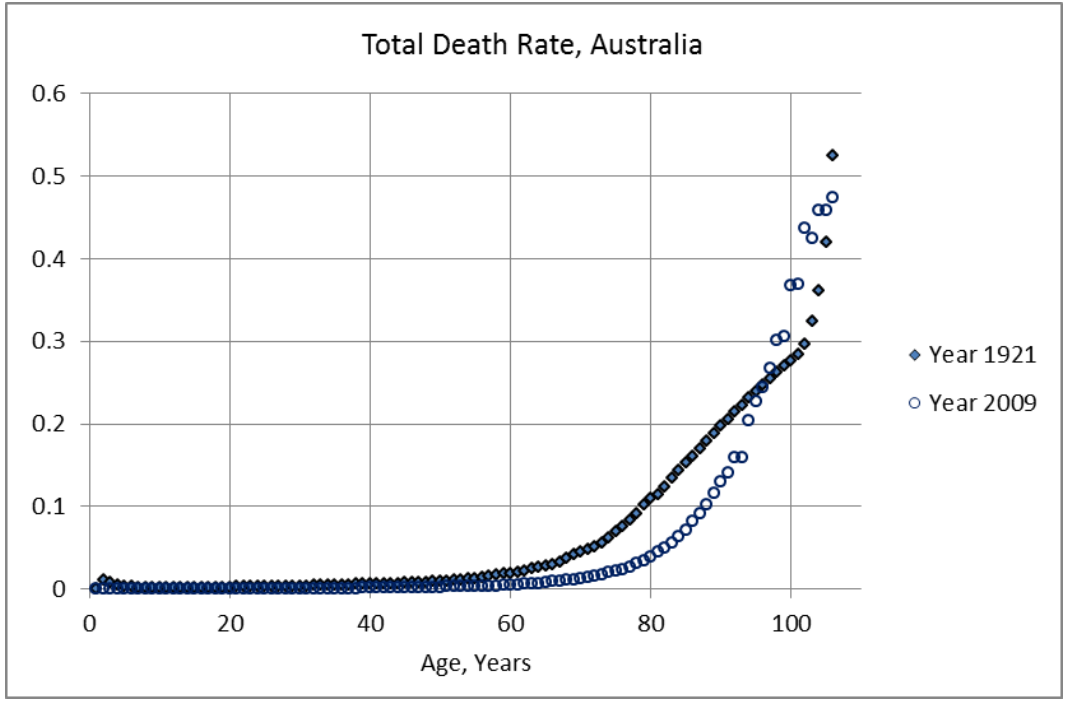

*Figure 2. Total (male and female) mortality rates in Australia in 1921 and 2009*

*Table 2. GTI and survival functions for Australia in 1921 and 2009*

| Cut-Off lifetime (age), T, Years | Australia, 1921 | | Australia, 2009 | |
|---|---|---|---|---|
| | GTI(T) | Survival Function $S(T)$ | GTI(T) | Survival Function $S(T)$ |
| 25 | -0.578 | 0.868 | 0.168 | 0.899 |
| 65 | 0.208 | 0.563 | 0.516 | 0.892 |
| 105 | 0.669 | 0.0005 | 0.797 | 0.002 |
| Median Age at Death | 68 | | 84 | |

Table 2 reveals that at the cut-off 65 and 105 years the ageing of Australian population increased. It is also interesting to note that at the cut-off 25 years the 1921 population was rejuvenating possibly due to a considerable infant mortality in 1921 compared to 2009 when it became ageing.

V. CONCLUSION

The study shows that the notions and statistics of ageing distributions used in mathematical theory of reliability, and the Gini-type index (GTI) specifically can be applied to investigation of human population ageing and longevity. In opposite to the median and mean lifetime, the GTI needs one year data only to reveal the population ageing or rejuvenating.

It should be noted that the suggested application of the GTI is not limited to investigating ageing and its trend in different countries and cohorts; it might be also applied to research of possible dependence between the macroeconomic Gini index and the GTI as measure of population ageing.